\documentclass[a4paper,11pt]{article}
\usepackage[total={6in,8in}]{geometry}
\usepackage[utf8]{inputenc}
\usepackage{authblk}
\usepackage{textcomp,amssymb,graphicx,epsf,cite}
\usepackage{hyperref}
\usepackage{epstopdf}
\usepackage{amsbsy}
\usepackage{amsmath}
\usepackage{mathtools}
\usepackage {xcolor}
\usepackage{graphicx,epsfig}
\usepackage{float}

\begin{document}
\title{Kinetic Exchange Models of Income and Wealth Distribution : Self Organization
and Poverty Level}
\author[1]{Sanjukta Paul}
\author[2, 3]{Bikas K. Chakrabarti}
\affil[1]{Department of Physics, Bankura University, Purandarpur, Bankura, West Bengal 722155, India}
\affil[2]{Saha Institute of Nuclear Physics, Kolkata 700064, India}
\affil[3]{Economic Research Unit, Indian Statistical Institute, Kolkata
700108, India
}
\date{\today}
\maketitle

\begin{section}{Introduction}
\label{Intro}
 Our thought process is often wired to rely heavily on 
a reductionist approach while trying to explain any observed complex physical phenomena.
This helps especially when we fail to understand the basic principles governing them.
What we essentially do is break up a physical event into the smallest possible snapshots of incidents, dismantle 
a machine to its constituent parts and tear apart an atom
to find its elementary particles, so that we know how the basic 
units actually behave. While reaching the roots 
of a physical problem can at times give us piecewise information, that can add up 
to generate the bigger picture, it may not always give a one-to-one correspondence
between the two. Thus, conjecturing the 
characteristics of the collective behavior by always studying the properties of the 
elements can be quite misleading! Here come the features of any many-body
theory, which differs considerably from the single-body behavior.
A simplistic guess could be that the additive laws our mental imagery
uses to yield a many-body picture from a collection of single bodies, 
can often give way to laws of multiplication which drastically
change the results. In the context of physics, this is termed \textquotedblleft emergence\textquotedblright,
which means giving rise to combined or cooperative properties that are starkly different from the individual characteristics. In the literary language, it simply translates 
to the child being very different from the parents!

The field of econophysics is one such emergent interdisciplinary area,
which briefly stated, tries to employ established
and tested laws of physics (in particular the many-body physics
or statistical physics of condensed matter systems) to model
and comprehend social behavior or, in particular, market
behavior. Both these social or economic systems are many-body
systems, though the number of basic constituents, $N$, (\textquotedblleft Social atoms\textquotedblright $\sim 10^2$ in case of rural markets 
and $\sim 10^9$ in case of global markets)
is much less than the Avogadro number $\sim 10^{23}$ (as in physical
condensed matter systems). The reason for the expected success
of econophysics comes from the observation that no established
law in one branch of natural science (say physics) becomes invalid
in another (say in biology or chemistry), though the importance
may not be uniform in all the branches of natural sciences.
Also,
physics has been the oldest of all the natural
sciences to get firmly established and so has already acquired a lot of armory over more
than three centuries. Consequently, the branches like physical
chemistry, chemical physics, biophysics, geophysics etc.
have 
developed and contributed significantly to the respective mother
disciplines. In a similar way, the recent attempts to develop interdisciplinary
fields like 
econophysics or sociophysics have followed quite inevitably.
Indeed, the early approaches to develop
econophysics (see e.g., Saha and Srivastava
\cite{Saha_Srivastava} in 1931, Majorana \cite{mantegna}
in 1942, Mandelbrot \cite{Mandelbrot}
in 1960, Mantegna and Stanley \cite{stsn-manteg} in
1995, and Chakrabarti and Marjit\cite {Marjit BKC} in 1995) proposed
precisely this natural path. Continuing this,
the recent attempts to develop econophysics
(see e.g., \cite{stanley,chakrabarti2006econophysics,sinha2010econophysics}
for reviews) and sociophysics
(see e.g., \cite{galam,sen2014sociophysics} for reviews)
have made considerable progress.

The kinetic theory has been the oldest (more
than a century old) many-body theory, starting
with the classical ideal gas model (see e.g.,
\cite {Saha_Srivastava}). The theory has undergone several
tests over ages and has made important analytical
progress over the years. It has therefore been a
instinctive choice of econophysicists for developing
the first-order models and science of markets,
and society in general. In particular, the model
studies on income or wealth distribution in
societies have received major attention and
obtained reasonable success (see e.g., \cite{yakovenko2009colloquium,chakrabarti2013econophysics,pareschi2013interacting,ribeiro2020income,greenberg,Economics_BKC}; see also \cite{Economics_BKC} arguing why econophysics, economics
for that matter, has to become an integral part of
natural science).

We will talk about Kinetic Exchange Models (KEMs) in detail later (see Sec. \ref{KEM}) after 
touching upon the concepts of Inequality indices (in Sec. \ref{II_Intro}), since their existence has predated these 
models in other fields of study and which play pivotal roles in understanding realistic aspects of the models. Discussions on the concept of Self-Organization and its relevance in KEMs follows
in Sec. \ref{SO}, with the summary in Sec. \ref{Summary} winding up the current chapter .
\end{section}

\begin{section}{Inequality Indices}
\label{II_Intro}
 Societies, since time immemorial, have been rife with inequalities
 and it is important to study them carefully since they have far-flung consequences
 like fluctuations in economy, political instability, civil wars
 and global conflicts. Inequalities can be broadly categorized into inequalities of conditions
 that imply inequality in the wealth, asset and income distributions and inequalities of 
 opportunities referring to unequal access to \textquotedblleft life chances\textquotedblright~ like education, health facilities and so on. Inequalities arise because
 human interactions, in the realm of trade, are quite complex and the resulting dynamics
 often lead to them quite stochastically. However, the inequalities can be easily computed (see for e.g. \cite{ghosh2014inequality,Entropy_BKC}) with relative measures of income 
 using certain indices, so that we can easily regulate them, rather than studying broader distributions of wealth or income,
 involving log-normals and power laws.
 
 \begin{figure}[!tbh]
  \centering
  \includegraphics[width=0.7\textwidth]{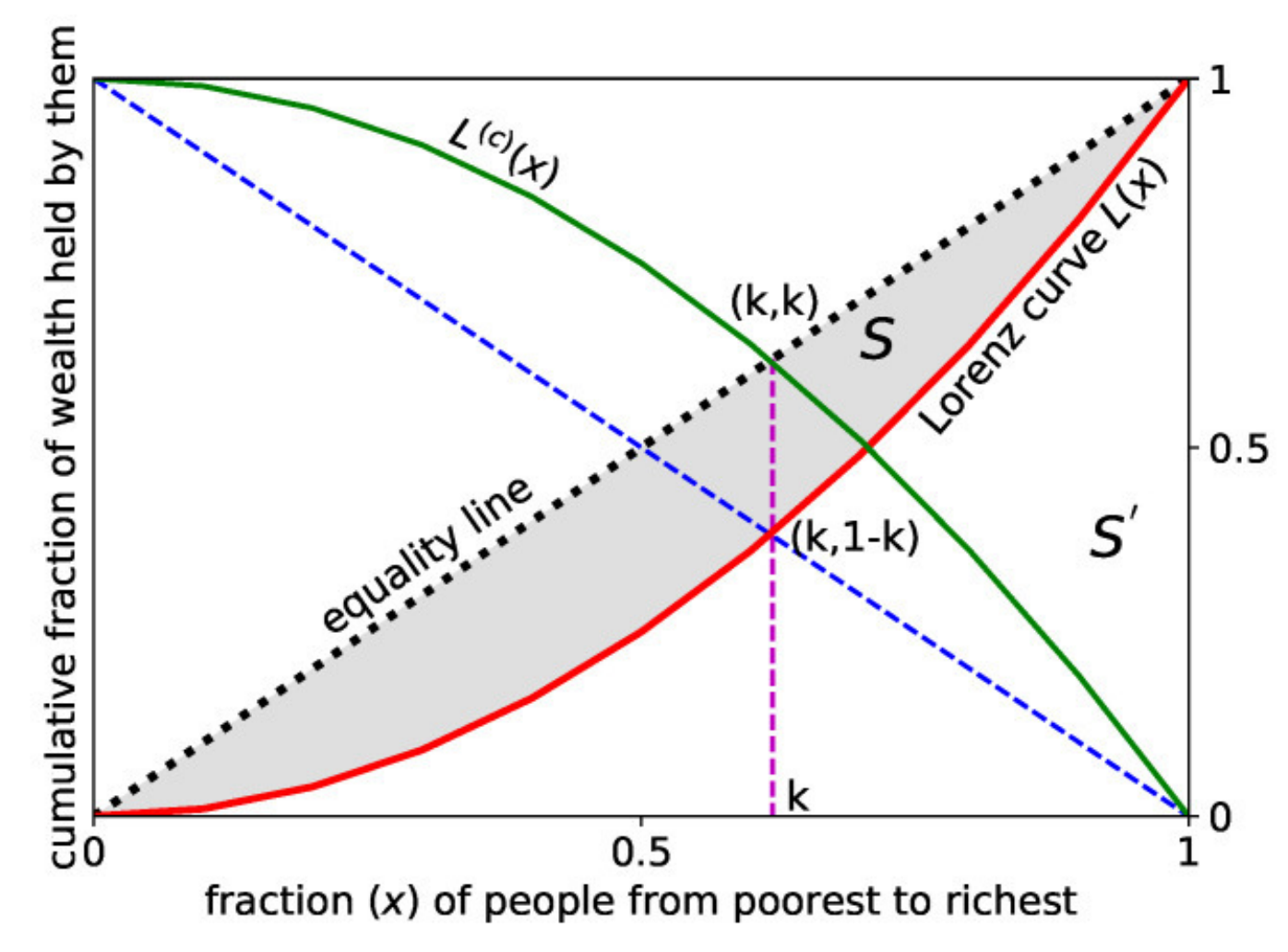}
  \caption{ $L(x)$, the Lorenz function, shown as a red curve, represents the
fraction of overall income or wealth assumed by the
 fraction $x$ of the people, when arranged from lowest to highest income.
 The Gini index $g$ is the ratio of the area
between the line of equality and Lorenz curve, normalized by 
the total area under the line of equality ($g = 2S$). The
complementary Lorenz function $L^c(x) \equiv 1-L(x)$,
given by the green curve, is essential for defining the $k$ index. 
The $k$ index is given by the ordinate value of the 
intersection of $L^c(k)$ with the diagonal perpendicular to the equality line.
It is a fixed point $L^c(k) = k$
of $L^c(x)$.}
    \label{fig:Lorenz_fnc}
\end{figure}
 \begin{subsection}{Gini Index ($g$)}
\label{II_Gini}
 The most traditional and commonly used inequality index is the Gini index $g$ (\cite{ghosh2014inequality,Entropy_BKC}). To define $g$, we need to remind ourselves of the Lorenz curve which represents the cumulative fraction $L(x)$ ($L = \int_{0}^{m} mP(m) dm/[\int_{0}^{\infty} mP(m) dm]$) of income or wealth accumulated (or citations of articles, cumulative proportion of species in a biodiversity, etc.) to the cumulative proportion $x$ ($\int_{0}^{m} P(m) dm/$ $[\int_{0}^{\infty} P(m) dm]$) of individuals (or entries) in a population, ordered from poorest to richest. 
We can thus readily infer that if every individual or trader
possessed equal wealth or income, then we would get the equality line $L(x) =x$ to be
passing through the origin (refer Fig. \ref{fig:Lorenz_fnc}). An inequality measure would immediately suggest that we consider the deviation
from this perfect equality line. Since we are considering a dispersion of the income or wealth
on a 2D plane, it is common to use the area enclosed by the straight line and the 
curve to represent the degree of inequality. However, it is imperative to use ratios of areas
to reduce the Gini index ($g$) to just a number so that $g= 0$ represents perfect equality, where every individual has the same money. On the other extreme, $g =1$ represents perfect inequality, where
the total money gets acquired by a single individual in the market.
$g$ is thus defined as the ratio of the area between the equality curve and the Lorenz curve $S$ to the area below the equality line $(S+S^\prime)$ (see Fig. \ref{fig:Lorenz_fnc}). A simple observation suggests
that $g$ is actually equal to $2S$. 
\end{subsection}

\begin{subsection}{Kolkata Index ($k$)}
\label{II_Kolkata}
A relatively newer inequality index than $g$, called the Kolkata index $k$, first proposed in \cite{ghosh2014inequality}
to account for acute social inequalities in Kolkata, is given by
the y-coordinate $k$ of the intersection point of the Lorenz curve
and the diagonal perpendicular to the equality line (see Fig. \ref{fig:Lorenz_fnc} and \cite{banerjee2020inequality} for some details). $k$ is the non-trivial
fixed point of the complementary Lorenz function $L^{c} (x) = 1 - L(x)
$, similar to $x = 0$ ,$1$ being the non-trivial fixed points of $L(x)$. Kolkata index states that fraction $(1 - k)$ of the population possesses $k$ fraction of the total 
wealth. This index essentially vouches for the validity of the 80 - 20 fractions 
of Pareto's law by indicating that fraction $(1 - k)$ of the population earns more money or possesses more wealth than fraction $k$ of the people (for some early report on observations and
data analysis, see \cite{chatterjee_rel_g_k}). The two limits of the 
$k$ index are $k = 1/2$ for $g = 0$ and $k =1$ for $g = 1$.
\end{subsection}

\end{section}

\begin{section}{Kinetic Exchange Models for Economy}
\label{KEM}
 The kinetic
theory of gases deals with the behavior of 
a large number of particles, by studying their interactions in the microscopic (small scale) detail 
to calculate macroscopic (large scale) parameters \cite{Saha_Srivastava}. Thermodynamics, however, has been the oldest discipline that establishes relationships between the macroscopic parameters in equilibrium,
using few postulates but without delving into the microscopic details.
Statistical mechanics, being the most advanced in this lineage, stems from these two fields and using the laws of mechanics,
makes some valid statements about the microscopic properties of the system 
in equilibrium and connects them to the macroscopic properties.

\begin{subsection}{Maxwell-Boltzmann and Gibbs distribution in an ideal gas}
 \label{MB_Gibbs}
 In the conventional kinetic theory for gas models,
the gas molecules or atoms, at equilibrium and at 
a particular temperature $T$ collide with one another and 
exchange kinetic energy (they are treated as free particles neglecting the
potential energy), keeping the total energy conserved 
in the process. The energy distribution function in this case 
can be seen to be an exponential of the form $f(\epsilon) = \exp({-\epsilon/{\Delta}})$ 
(from the condition $f(\epsilon_1)f(\epsilon_2) = f(\epsilon_1) + f(\epsilon_2)$, 
$\epsilon_1$ and $\epsilon_2$ being energies of any two particles respectively) \cite{Saha_Srivastava},
where $\epsilon$ denotes the energy of a gas particle and $\Delta$ can be 
estimated from the equation of state of the gas. It must be noted that the gas particles
have a chaotic motion and that the change in velocity (or energy) of the particles 
after a collision is fully random. This implies that at an instant 
of time, even though particles may have the same velocity or energy, after a collision
process some have these quantities increased and the others reduced.
From the distribution function, we can easily calculate the 
number density of gas particles \cite{Saha_Srivastava} having energy $\epsilon$
to be $n(\epsilon) = \sqrt{\epsilon} \exp({-\epsilon/{KT}})$,
where $\Delta = KT$. The factor 
$\sqrt{\epsilon}$ ensues from the density of states (in 3 dimensions) \cite{Saha_Srivastava},
which when multiplied to the distribution function $f(\epsilon)$, gives rise 
to the number density $n(\epsilon)$. This gives the Gamma Distribution or the 
Maxwell-Boltzmann Distribution, where a peak or most probable
energy corresponding to a non-zero energy possessed by 
the largest number of gas particles, exists. Particles with very low and very high energy
are considerably less in number, following energy the distribution function (see \cite{Entropy_BKC}
for an elaborate discussion).
\end{subsection}
\begin{subsection}{Gibbs Distribution in Kinetic Exchange Models of market}
\label{Gibbs_in_KEM}
 Translating the kinetic theory for gases
to the domain of market models
in an economy (see for e.g. \cite{Dragulescu Yakovenko 2000,chakraborti2000statistical,chatterjee2004pareto,chatterjee2007kinetic}), we can find equivalences and differences between the two.
In the KEMs for wealth distribution, we have 
random traders or agents that act as socially interacting atoms 
taking part in a trade, where random exchanges of wealth
is conducted, much like the collisions between the gas particles 
resulting in exchange of energies. Wealth, in general, is 
usually considered to be the solid assets, one individual or an organization 
possesses, having an economic value. 
However, it can broadly represent the fortune or well being 
of a particular individual. In these models, wealth denotes the exchange of 
a more fluid entity called money, that traders use for bartering 
intangible services or tangible objects of interest. It thus 
is an analogy for economic energy and is associated with 
an economic temperature $\Delta^{\prime}$, a constant for a 
trading process. Since there is no counterpart of the particle 
momenta in the trade market, the density of states is a constant and hence
as before (see Sec. \ref{MB_Gibbs}), the number density $n(m)= c f(m)$, where 
$f(m)\sim\exp(-m/{\Delta^\prime})$ (with the condition that $f(m_1)f(m_2) = f(m_1) + f(m_2)$, $m_1$ and $m_2$ being wealth of any two agents respectively) is the money distribution function.
We note that the total 
number of traders $\int_{0}^{M} n(m) dm = N$,
is a constant in a market, although the buyer of 
one trading transaction may be the seller of the other 
and vice versa. The total money is $\int_{0}^{M} mn(m) dm = M$
and without considering debt in the system, it is a constant for a market.
Therefore, $\Delta^{\prime}$, identified as $M/N$ (the average money per person) is a constant.
Thus we conclude that if the traded amount is a fraction of the
wealth possessed by an agent, subject to the 
conservation of $M$, the resulting distribution at equilibrium 
is undoubtedly the Gibbs or exponential distribution. An exponential 
distribution immediately points at the largest number of paupers in 
the system to be the condition for the maxima in the distribution, indicating 
an unequal distribution of wealth, where most of the money
gets concentrated in a few rich hands. This inequality
strongly underpins the absence of an equivalent equipartition
theorem, applicable for gas models, that renders equal kinetic 
energies of $\frac{1}{2} KT$ for each degree of freedom 
in a polyatomic gas molecule. The equipartition law, stems from the 
ability to cast the kinetic energy of a gas molecule in terms of squares
of momenta, which has no analogy in the market model\cite{Entropy_BKC}.
\end{subsection}

Now, we present from \cite{yakovenko2009colloquium}, 
a purely observational income distribution data in Fig. \ref{fig:income_Yakovenko} to 
illuminate the nature of the different kinds of distribution 
elaborated in Secs. \ref{MB_Gibbs} and \ref{Gibbs_in_KEM}.
This plot beautifully demarcates the lowest income group exponential distribution 
from the highest income section of the society, following a power law decay (Pareto tail).
\begin{subsection}{Estimates of $g$ and $k$ for the Gibbs distribution}
\label{g_k_Estimate}
{To understand $g$ and $k$ quantitatively, 
we can evaluate them for the simplest case where 
P(m) is the normalized Gibb's distribution, $P(m) = \exp(-m)$.
Now as,
$ x =\int_0^m \exp (-m')dm' = 1- \exp(-m)$, it yields
$ m = - \ln(1-x)$, and hence $ L = \int_0^m m'\exp(-m')dm'$
= $ 1 - (m + 1)\exp(-m)$. Thus, $ L(x) = $
$ 1 - (1 - x)[1- \ln(1 - x)]$. Since, the area under the
equality line is the area of one of the two triangles with equal areas,
so it is 1/2. The corresponding $g = $
$ 1-2\int_0^1 L(x)dx = 1/2$, $L(x)$ being the Lorenz curve. The Kolkata index $k$
for this distribution is therefore given
by the self-consistent equation $ 1 - k = L(k)$ or
$ 1 - 2k = (1-k)[\ln(1 - k)]$, that produces $ k \simeq 0.68$.}
\end{subsection}
\begin{figure}[!tbh]
\includegraphics[angle=-90,width=0.95\linewidth]{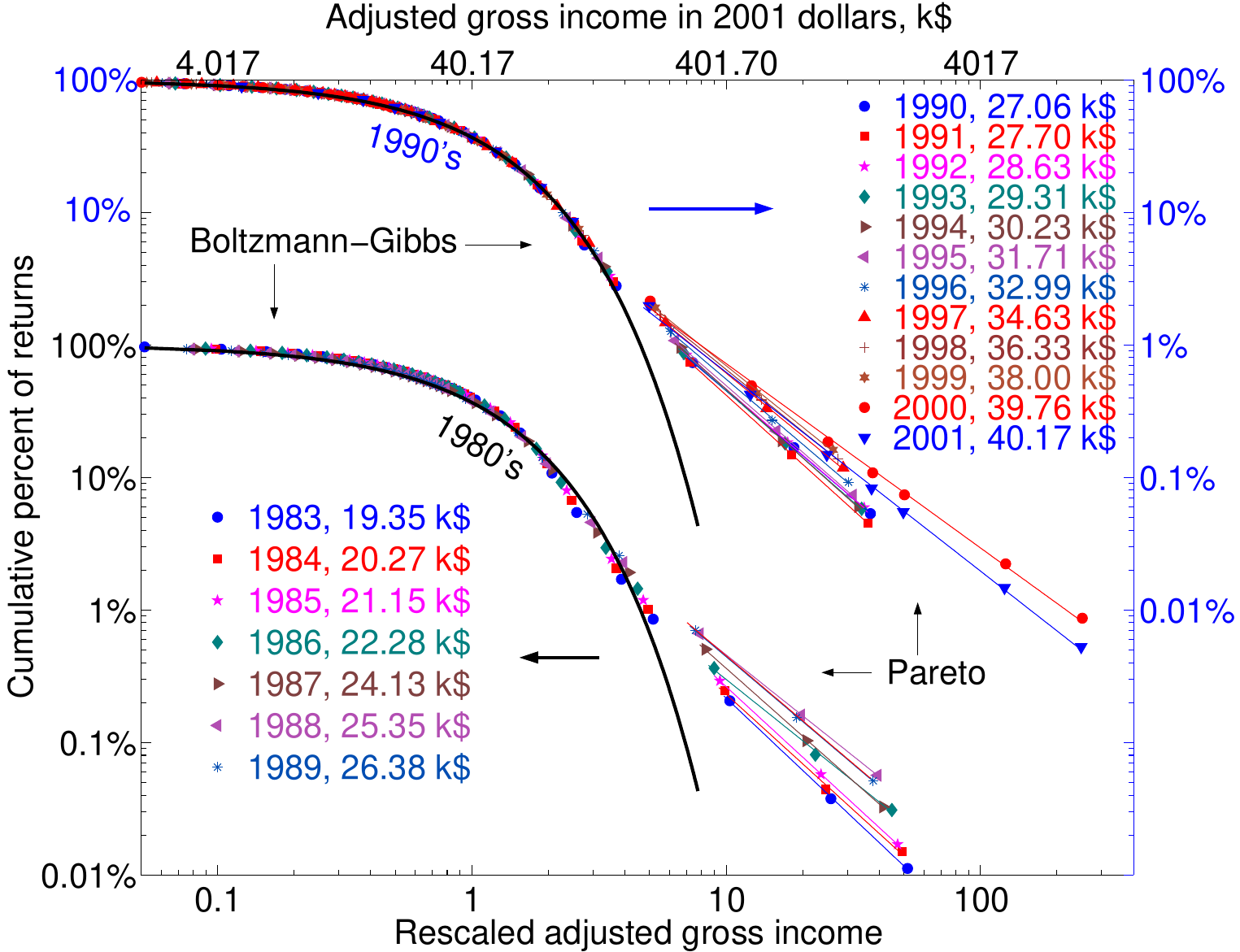}
\caption{The US income distribution data
(obtained from income tax returns) collapse
for the period 1983\textendash2001. The cumulative
percentage of returns has been plotted against
normalized income in log-log scale.
The fitting values of the effective
temperature (obtained by fitting the data
for the exponential part of the distribution;
see Sec. \ref{No_lambda}) have been given against each
year. This figure has been reproduced from
\cite{yakovenko2009colloquium} (permission from American Physical
Society).}
\label{fig:income_Yakovenko}
\end{figure}
\begin{subsection}{Basic computational framework for Kinetic Exchange Models}
\label{Comp_framework}
 Two agent interactions in a closed economy, with total number of agents
$N$ and total money $M$ can be considered analogous to a closed thermodynamic system
with total energy and total number of gas particles as constants. Any two randomly selected agents can carry out trade among themselves following a Markovian dynamics, where exchange 
of money at any time $t+1$ does not depend on any other exchanges other than $t$. The exchange mechanism $\mathcal{M}$ can be defined as
\begin{eqnarray}
 \begin{bmatrix} m_{i}(t+1)\\ m_{j}(t+1) \end{bmatrix}
 =
 \mathcal{M}
  \begin{bmatrix}
    m_{i}(t)
   \\
    m_{j}(t)
   \end{bmatrix}
\label{Exch_M}
\end{eqnarray}
Here, each agent or trader possesses money or wealth $ m_i(t)$ at time $t$ and any two random traders $i$ and $j$ exchange money by participating in the trading process, with no provision for debt ($ m_i(t)\geq 0$). Since conservation of total money is maintained in the whole
trading process, we have 
\begin{eqnarray}
 m_i(t+1) = m_i(t) + \Delta m\nonumber\\
 m_j(t+1) = m_j(t) - \Delta m
 \label{Exch_with_delta_m}
\end{eqnarray}
and 
\begin{eqnarray}
 m_i(t) +m_j(t) = m_i(t+1) +m_j(t+1) 
\end{eqnarray}
$\Delta m$ represents the fraction of money that is used for trade.
Each trade corresponds to an increase by an unit time interval (see \cite{paul2022kineticexchange} for
more realistic saving behaviors).
\end{subsection}
\begin{subsection}{Without-saving income exchange models}
\label{No_lambda}
 It is more realistic to use a fraction of the money a trader possesses,
 rather than invest the whole of it in the interaction. If $0 < \epsilon_{ij} < 1$
 represents a uniform stochastic fraction, with an average precisely equal to 0.5,
 \begin{eqnarray}
  \Delta m = \epsilon_{ij}[m_i(t) + m_j(t)] - m_i(t)
  \label{Without_lambda}
 \end{eqnarray}
The steady state always gives the Gibbs Distribution, $ P(m) = 1/T \exp(-m/T),$
$T=M/N$,
where it is immediately obvious that most of the population has very less money corresponding to the
exponentially decaying curve, irrespective of whether the initial 
distribution was uniform (see \cite{chatterjee2007kinetic} for a detailed 
description).
\end{subsection}

\begin{subsection}{Uniform-saving income exchange models with inequality indices}
\label{Uniform_lambda}
 Although the conservation of total money is a necessary condition for the 
 above models to work as reliable econophysical models, the number of paupers 
 is too large for them to point towards a sustainable economy. By this we mean that
 a trader is prone to lose all his money to a fellow trader 
 in a single trade since the exchanges are completely random,
 keeping the total money conserved. This is in synchronism to the kinetic theory of gases, where one gas particle can lose the entire energy 
 in a single collision.
 This brings us to a more realistic structure of these exchange models, where
 savings appears quite organically for individuals unlike gas molecules.
 If a saving propensity is introduced, traders will save a fraction of their money 
 before involving the rest of the money in trading and hence will avoid being paupers.
 A trader can only become a pauper if he loses in every trade 
 and gradually loses the saved fraction of the money, which is not possible because
 exchanges are random. Thus, for any non-vanishing saving propensity that remains uniform
 over time, the steady state distribution transitions from an exponential or Gibbs to a Gamma distribution, where we have zero number density of paupers.\\
 Let $\lambda$ be the uniform or fixed saving propensity of the traders. The exchange of money between two randomly chosen pairs $i$ and $j$ can be expressed as
\begin{eqnarray}
     m_i(t+1) &= \lambda m_i(t)+\epsilon_{ij}((1-\lambda)(m_i(t)+m_j(t))) \nonumber\\
     m_j(t+1) &= \lambda m_j(t)+(1-\epsilon_{ij})((1-\lambda)(m_i(t)+m_j(t)))
    \label{With_uni_lambda}
 \end{eqnarray}   
where $ 0 \leq \epsilon_{ij} \leq 1 $ is a random fraction varying in every interaction.

 The important aspect to note in Fig. \ref{fig:Uniform_lambda} here is that 
 introduction of a saving propensity immediately introduces a most probable income
 (MPI) in the distribution and we have the MPI shifting from $ m = 0$ for $ \lambda = 0$
 to $ m = M/N$ for $\lambda \rightarrow 1$. Thus, the Markovian dynamics 
 slowly gives way to a self-organizing feature of the whole market, even with the
 saving fractions on individual levels. The market thus essentially turns interacting for 
 non-zero $ \lambda$, with a global order and a critical behavior emerging, very different from the non-cooperative nature of the $ \lambda = 0$ limit.
 A natural extension of this would be the inclusion of a \textquotedblleft distributed-saving propensity\textquotedblright~, which has been studied in \cite{chatterjee2007kinetic}. For non-uniform saving propensity, the Gamma distribution collapses to a power law decay called the Pareto tail (see Fig. \ref{fig:income_Yakovenko}). The saving propensities readily generate the cross-over point for the Pareto distribution. 
 \begin{figure}[!tbh]
    \centering
    \includegraphics[width=1.0\textwidth]{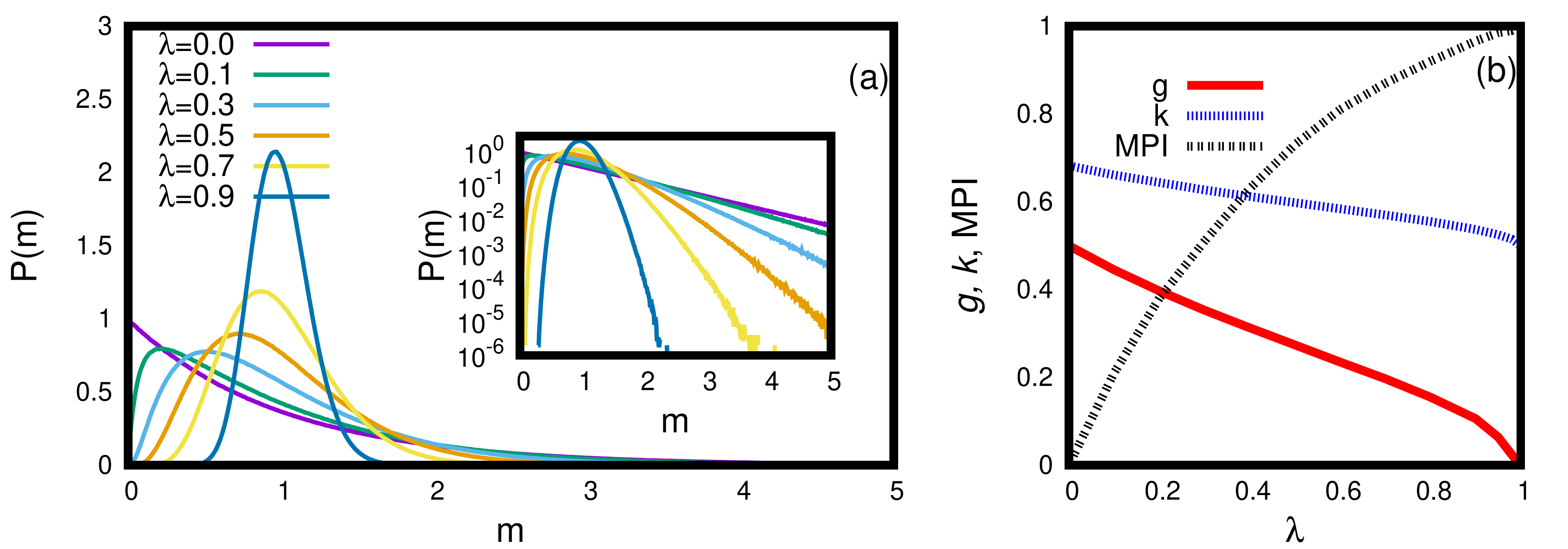}
    \caption{Kinetic exchange dynamics with uniform saving propensity :
    (a) $ P(m)$, the steady state income distribution, for fixed saving propensity $ \lambda$ has been plotted. The exponential nature of the tail end of the distributions is presented in semi-log scale in the inset. (b) The variations of the Kolkata index ($ k$), the Gini index ($g$) and location of the Self-Organized Poverty Level (SOPL) are plotted against fixed saving propensity $ \lambda$ (maximum value of $\rm \lambda = 1_{-}$).}
    \label{fig:Uniform_lambda}
\end{figure}
\end{subsection}
\begin{subsection}{Indices for pure kinetic exchange model with two annealed choices of 
$\lambda$}
\label{2_lambdas}
\begin{figure}[!tbh]
    \centering
    \includegraphics[width=1.0\textwidth]{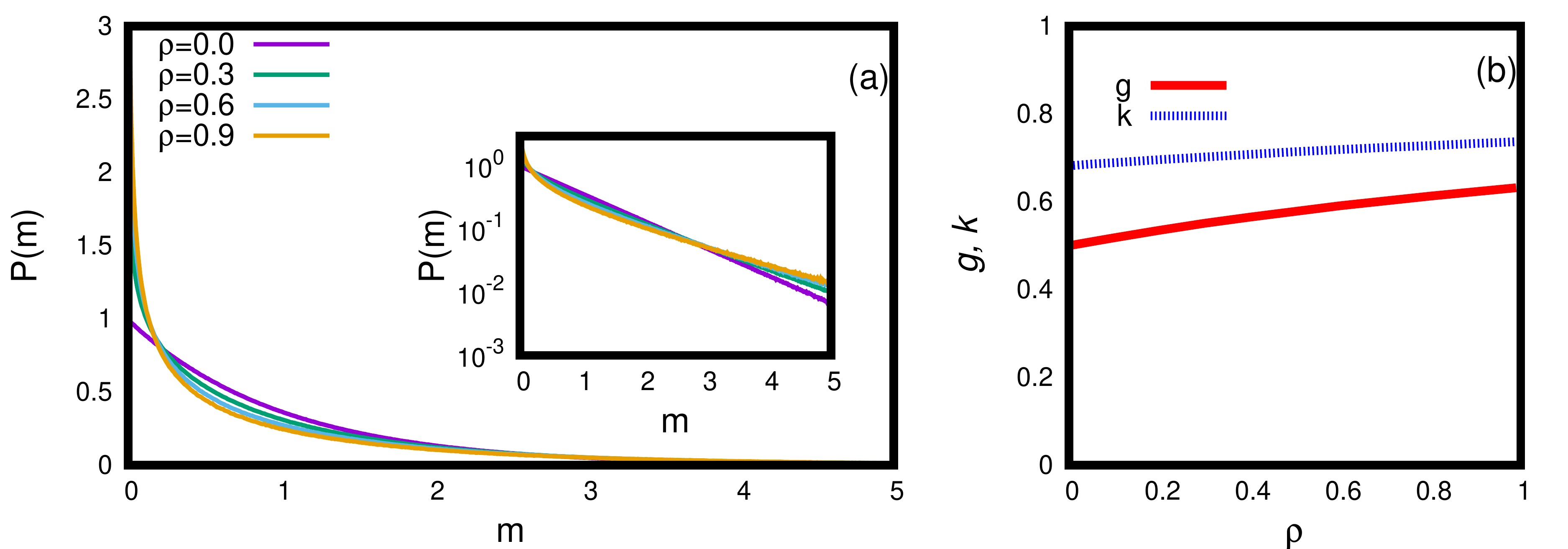}
    \caption{The pure KEM for two choices of $\lambda$.
    (a) $ P(m)$, the steady state income distribution, for fixed saving propensity $ \lambda$ has been plotted. The exponential nature of the tail end of the distributions is presented in semi-log scale in the inset. (b) The variations of the Kolkata index ($ k$), the Gini index ($g$) and location of the Self-Organized Poverty Level (SOPL) are plotted against the probability for taking $\lambda=1$ i.e. $\rho$ (maximum value of $\rho = 1_{-}$).}
    \label{fig:t_var_2_lambdas}
\end{figure}

In order to incorporate the inherent volatility of human saving behavior 
or take into account impalpable circumstances that might affect the constancy of the saving behavior, it would be wise to include a time dependence of $\lambda$. Hence, mathematically, treating the $\lambda$ probabilistically should apparently solve the problem. The multiplicative probability factor $\rm \rho$ thus determines the \textquotedblleft luck \textquotedblright for the saving behavior since we consider only the two extreme 
conditions of $\lambda_{\rho} = 1$ with probability $\rho$ and $\lambda_{\rho} = 0$ with probability $1-\rho$; $\rho <1$. 
Therefore, the slightly modified (from Eq. \ref{With_uni_lambda}) 
exchange dynamics goes as follows:\\
 \begin{eqnarray}
  m_i (t + 1) = \lambda_{\rho} m_i (t) +\epsilon_{ij} ((1 - \lambda_{\rho})(m_i (t) + m_j (t)))\nonumber\\
  m_j (t + 1) = \lambda_{\rho} m_j (t) + (1 - \epsilon_{ij} )((1 - \lambda_{\rho})(m_i (t) + m_j (t)))
  \label{Time_var_lambda}
 \end{eqnarray}\noindent {
It is to be noted that $\lambda_{\rho}$ will be treated as $\lambda$ from now on,
provided Eqn. \ref{Time_var_lambda} is satisfied.}

Eqn. \ref{Time_var_lambda} suggests that each trader has two choices of $\lambda$ over time.
We observe in Fig. \ref{fig:t_var_2_lambdas} that the money distribution regains its exponential nature pushing the MPI to zero again and the inequality indices $g$ and $k$ increase minimally with $\rho$.
 It can be readily inferred that although a probabilistic factor has been 
introduced into the picture, $\lambda$ can only oscillate between two distinct and extreme values of 1 (with probability $\rho$) and 0 (with probability $1- \rho$) . It may be noted that when $\lambda = 0 $, we have the well known 
Gamma Distribution (see Fig. \ref{fig:Uniform_lambda}) where the MPI is close to zero and for $\lambda = 1$, the exchange process stops since money is only saved and not traded. Realistically, for $ \lambda \sim 1$ (that has been studied for simulations in \cite{paul2022kineticexchange}), very few people save the total money in the system and most of the population go penniless. Although, the time-averaged distribution may look similar for both the limits, the number of people being paupers is far higher for $\lambda \sim 1$. We thus see in Fig. \ref{fig:t_var_2_lambdas} that for $\rho = 0$ (which implies $\lambda = 0$ with probability 1) we regain the $\lambda = 0$ exponential of Fig. \ref{fig:Uniform_lambda}. Also, for higher $\rho \sim 1$, we have a very high probability of having $\lambda \sim 1$, leading to a steeper exponential. Therefore, there can be no intermediate $\rho$
value possible for which we get a non-zero MPI and the inequality indices ($g$ and $k$) only increase with $\rho$ (Fig. \ref{fig:t_var_2_lambdas}).

\end{subsection}
\end{section}

\begin{section}{Self-Organization}
\label{SO}
The concept of self-organization is not unique to the field of market models
but to various other fields where \textquotedblleft avalanches\textquotedblright~ occur due to dynamical interactions
of individual particles, without the help of any external disturbance.
In the market models, it can be readily inferred from Sec. \ref{2_lambdas} and Fig. \ref{fig:t_var_2_lambdas} that
there should be a bias in the selection of agents,
so that the non-zero MPI condition can be revived (for important observations, see \cite{Iglesias,pianegonda2003wealth,ghosh2011threshold}). In \cite{paul2022kineticexchange},
this has been explored by taking one of the agents to be the poorest of all 
in the trade and keeping the saving propensity time-independent and time-dependent
respectively (see Figs. \ref{fig:3} and \ref{fig:Time_var_lambda_with_poorest_agent} respectively). There appears a self-organized threshold value of money,
with no trader having money less than that, arising purely due to the 
spontaneous interaction between the agents called the Self-Organized Poverty Level (SOPL).

\begin{figure}[!tbh]
    \centering
    \includegraphics[width=1.0\textwidth]{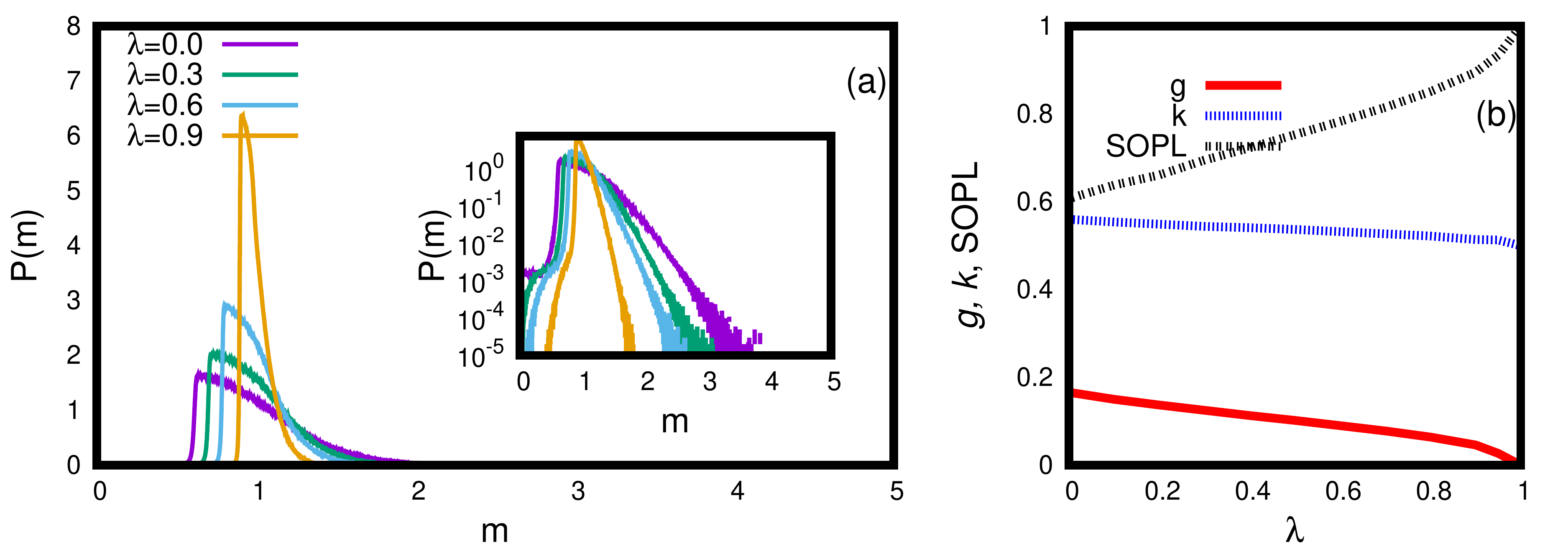}
    \caption{Self-Organized Poverty Level model: (a) $ P(m)$, the steady state income distribution, for fixed saving propensity $ \lambda$ has been plotted. The exponential nature of the tail end of the distributions is presented in semi-log scale in the inset. (b) The variations of the Kolkata index ($ k$), the Gini index ($g$) and location of the Self-Organized Poverty Level (SOPL) are plotted against fixed saving propensity $ \lambda$ (maximum value of $\rm \lambda = 1_{-}$).}
    \label{fig:3}
\end{figure}

\begin{subsection}{Self-organized poverty level in Kinetic Exchange models and inequality indices}
\label{SO_in_KEM}
For a time-independent $\lambda$, one of the agents is favorably chosen, provided the agent has the lowest money while the other agent is randomly chosen from the rest of the population. The money exchange mechanism is the same as in Eqn. \ref{Time_var_lambda}. It has been observed in \cite{ghosh2011threshold,paul2022kineticexchange} that not only does the MPI shift to the right end of the money distribution (close to $m =1$ for larger values of $\lambda$) but also an SOPL appears (as in Figs. \ref{fig:3} and \ref{fig:Time_var_lambda_with_poorest_agent}), even for $\lambda = 0$, that was previously absent. 
The SOPL has been found to increase with the increase in $\lambda$ (see Fig. \ref{fig:3}).
It is to be emphasized that the inequality indices $g$ and $k$ are also seen to
decrease with $\lambda = 0$ (Fig. \ref{fig:3}), establishing the fact that the mere advantage of 
being afloat in the trading market on a local scale, without losing the ability to trade 
because of recurrent losses in the exchange, brings about a collective
change in the global economy.

\end{subsection}
\begin{subsection}{An estimate of the SOPL for $\lambda = 0$}
\label{SOPL_estimate}
We denote the SOPL (for saving propensity
$\lambda = 0$) by $\delta$ and assume the
money or wealth distribution $P(m)$
remains exponential ($\rm \exp(-m)$), with average
money = $M/N = 1$
(see Fig. 3 of \cite{paul2022kineticexchange}). Assuming a random (50 - 50
sharing) exchange between  the agent at money
$\delta$ and another with the average money
$(\delta +1)$ with probability $\exp (-1)$,
one gets
$\delta = (1/2)[ \delta + (1 + \delta)\exp(-1)$,
giving $\delta = \exp(-1)/(1 - \exp(-1))$, or
$\delta \simeq 0.58$, compared to numerically
estimated value (\cite{ghosh2011threshold,paul2022kineticexchange}) of $\delta \simeq 0.60$.
\end{subsection}
\begin{subsection}{Self-organized minimum poverty level model: Indices for two choices of $\lambda$}
\label{SOPL_2_lambdas}
\begin{figure}[!tbh]
    \centering
    \includegraphics[width=1.0\textwidth]{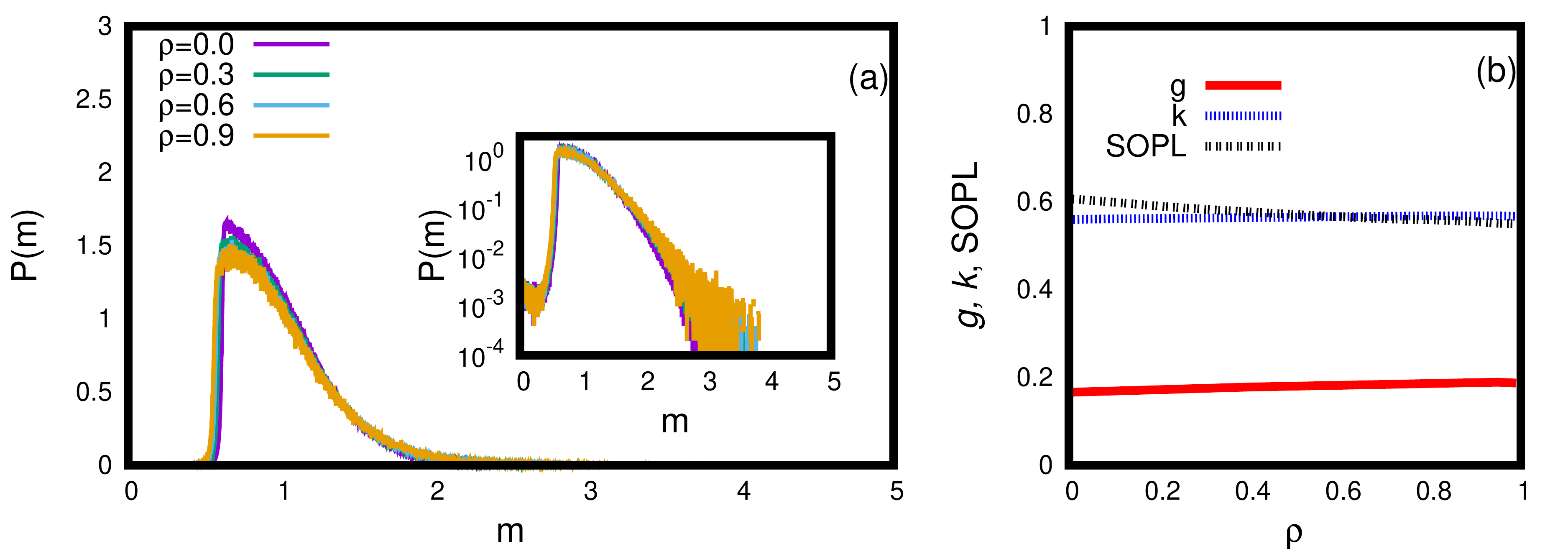}
    \caption{Emergence of a self-organized minimum poverty level model is shown for two choices of saving propensity: (a) $ P(m)$, the steady state income distribution, for fixed saving propensity $ \lambda$ has been plotted. The exponential nature of the tail end of the distributions is presented in semi-log scale in the inset. (b) The variations of the Kolkata index ($k$), the Gini index ($g$) and the Self-Organized Poverty-Line or SOPL are plotted against fixed saving propensity $ \lambda$. The inequality indices gradually increase along with the SOPL as a function of $ \rho$ (maximum value of $ \rho = 1_{-}$).}
    \label{fig:Time_var_lambda_with_poorest_agent}
\end{figure}

In this section we revisit the final case investigated in \cite{paul2022kineticexchange}.
$\lambda$, as before in Sec. \ref{2_lambdas} can take two values of 1 or 0, depending on the probability 
$\rho$ or ($1-\rho$). However, the biasness of always selecting one of the agents to be the poorest of all 
in the population remains fixed. The steady state and averaged 
distribution $P(m)$ varying with $m$ is provided in Fig. \ref{fig:Time_var_lambda_with_poorest_agent}. The remarkable change noted here is that
the time dependence does in no way kill the finite MPI and the SOPL formation (unlike in Fig. \ref{fig:t_var_2_lambdas}) ! Both are rather 
retained, making them robust with change in $\lambda$. However, the inequality 
indices $g$ and $k$ increase very slowly with $\rho$ and the SOPL reduces gradually with $\rho$.  
\end{subsection}
\end{section}

\begin{section}{Summary and Discussions}
\label{Summary}
In summary, we have invited the reader to engage in a broader 
discussion about the captivating field of econophysics (see Sec. \ref{Intro}) and how,
sitting on the fertile confluence of physics and economics, it seeks
quantitative answers about the questions of inequalities in economies, 
spontaneous re-organization of money distributions and triggers research 
on methods to improve the overall quality of economies in general.
We have tried to provide a nuance of the KEMs (refer Sec. \ref{KEM})
and the relevant inequality indices (see Sec. \ref{II_Intro}) needed to model any many-body economical market, using basic laws in physics. We have also extended our ideas on how they can be tuned to generate 
self-organizing features in the wealth distribution of a population, leading to 
a minimum poverty line and reduced inequality measures.
Since we have visualized economies through the lens of physics,
governed by precise laws that have been tested, we 
have more freedom to introduce certain constraints or biases
in the system that can mimic real world policies framed for the 
reduction of inequalities in income or wealth distributions
among people. We have thus enumerated various cases (mostly numerically; see for e.g.
\cite{paul2022kineticexchange,sinha2020econophysics}) 
where the basic KEMs have been manipulated (see Secs. \ref{KEM}, \ref{SO})
to include a saving behavior locally, with an impact that would 
grow manifold and show up in the global MPI; favor the poorest in the trading market,
which in the real world could mean giving them incentives;
or render the saving behavior time-dependent, which 
in real economies too could put a cap on the 
reduction of the inequality in possession of wealth.

While the world economy, in recent times, is abuzz with reports on condensation of
almost half the population's total wealth in the hands of a countable few rich,
it is essential to investigate whether the KEMs can capture such
observations. In the acclaimed Chakraborti model \cite{Chakraborti2002YSmodel} or the so-called
Yard Sale model, where for each trade the richer one offers the same amount as that
possessed by the other, the market dynamics stops when the entire money ends
up in the hands of one trader (and no further trade is possible with the paupers).
The KEM of Goswami and Sen \cite{Goswami_Sen_model},
where the trading probability between any pair of traders decreases 
with their wealth difference (following an inverse power of the wealth
difference between the two traders), indicated considerable wealth
concentration towards the richer. Following this indication, the Banerjee
model was proposed \cite{Banerjee_model}, where the trade
between just nearest neighbors in wealth at any stage showed \cite{asimghosh2023} that about $99.8\%$ of the total wealth gets accumulated in
the hands of exactly 10 traders (fortunes of these traders are not everlasting) in
the steady state for the macroscopic limit of the market size $N$. The novelty
of such KEMs (\cite{Chakraborti2002YSmodel,Goswami_Sen_model,Banerjee_model})
for these quasi-oligarchic societies is provided by the observation that the
presence of a non-vanishing probability of random exchanges can destroy the
condensation phenomena and can lead to exponential wealth distributions in the
large $N$ limit (see \cite{asimghosh2023}).
\end{section}

\end{document}